\documentclass{emulateapj}

\usepackage{amsmath}
\usepackage{natbib}
\usepackage[colorlinks=true,citecolor=blue,urlcolor=magenta,breaklinks]{hyperref}
\usepackage{cool}
\usepackage{eulervm}
\usepackage{times}

\begin{document}
\title[Gravitational Waves from Stellar Black Hole Binaries and the Impact on Nearby Sun-like stars]{ 
Gravitational Waves from Stellar Black Hole Binaries \\ and the Impact on Nearby Sun-like stars}
\author{Il\'\i dio Lopes~\altaffilmark{1}, Joseph Silk~\altaffilmark{2,3}}
\altaffiltext{1}{Centro Multidisciplinar de Astrof\'{\i}sica, Instituto Superior T\'ecnico, 
Universidade de Lisboa, Av. Rovisco Pais, 1049-001 Lisboa, Portugal; ilidio.lopes@tecnico.ulisboa.pt} 
\altaffiltext{2}{Institut d'Astrophysique de Paris, UMR 7095 CNRS, Universit\'e Pierre et Marie Curie, 
98 bis Boulevard Arago, Paris F-75014, France; silk@astro.ox.ac.uk} 
\altaffiltext{3}{ Department of Physics and Astronomy, 3701 San Martin Drive, 
The Johns Hopkins University, Baltimore MD 21218, USA }


\begin{abstract} 
We investigate the impact of resonant gravitational waves on quadrupole acoustic modes of Sun-like stars
located nearby stellar black hole binary systems (such as GW150914 and GW151226). We find that the stimulation
of the low-overtone modes by gravitational radiation can lead to sizeable photometric amplitude variations,
much larger than the predictions for amplitudes driven by turbulent convection, which in turn are consistent
with the photometric amplitudes observed in most Sun-like stars.
For accurate stellar evolution models,
using up-to-date stellar physics, we predict photometric amplitude variations of $1$ -- $10^3$ ppm for a solar
mass star located at a distance between 1 au and 10 au from the black hole binary, and belonging to
the same multi-star system. The observation of such a phenomenon will be within the reach of the {\sc Plato}
mission because telescope will observe several portions of the Milky Way, many of which are regions of high
stellar density with a substantial mixed population of Sun-like stars and black hole binaries.
\end{abstract}

\keywords{
Stars: general -- black hole physics -- asteroseismology -- binaries: general -- globular clusters: general -- X-rays: binaries
}

  
\section{Introduction}

The ground-breaking observations of gravitational waves
from the coalescence of two black hole binaries~\citep{2016PhRvL.116f1102A,2016PhRvL.116x1103A}, 
swiftly initiated the age of gravitational-wave astronomy.
These discoveries suggest that massive stellar black
holes are  more common than previously thought. As a
consequence, the idea that intrinsic oscillations of stars may
be stimulated by gravitational radiation from nearby stellar
black hole binaries becomes a real possibility~\cite[i.e.,][]{2015ApJ...807..135L}.

\smallskip
A compelling reason for our study comes from the fact that
stellar oscillations stimulated by nearby stellar black hole binaries
might have already been observed unintentionally by
current missions. At the very least, such observations are now
within the reach of future asteroseismology missions. During
the last decade, 
the {\sc Corot}~\citep[launched 2006,][]{2006ESASP1306...33B}  and {\sc Kepler}~\citep[launched 2009,][]{2010ApJ...713L.160G,2014PASP..126..398H}
satellites, despite their restricted fields of view, have discovered
non-radial oscillations in approximately 700 main-sequence
and sub-giant stars and more than 17,000 red-giant stars out of
the 400 billion stars expected to exist in the 
Milky Way~\citep[i.e.,][]{2012EPJWC..1905012M,2013ARA&A..51..353C,2015ApJ...809L...3S}.
In these stars, known as Sun-like stars, such non-radial oscillations
are driven stochastically by the turbulence of their
external convective envelope. The {\sc Kepler} mission alone has
measured oscillations in stars located at distances up to 15
kpc, almost two times the distance of the Sun to the Galactic
center. 
This mission has observed such stars uninterruptedly through long periods of time, typically with a duration varying from 3.5 to 4 years. In many of the cases, astronomers were able to determine the high-precision spectrum of their stellar non-radial oscillations.
Such stars are located in contrasting space
environments of the Milky Way, within the galactic disk
and bulge, and in halo globular clusters. Many of these
locations are dense stellar environments, such as globular
clusters with mixed populations of Sun-like stars and black hole
binaries~\citep{2015PhRvL.115e1101R}. Therefore, it is to be
expected that some of these Sun-like stars are located near
these binaries and close enough for the incoming
gravitational waves to excite some of their non-radial modes~\citep{2015ApJ...807..135L}. 
In agreement with our previous work,
such Sun-like stars are referred to as  star  detectors.

\smallskip
Most likely the Sun-like star and the stellar black hole
binary will be part of the same gravitationally bound
multi-body system. There is theoretical and observational
evidence pointing to the formation of multi-star
systems~\citep{2015ApJ...807..135L}. Furthermore, the current
theory of star formation predicts the existence of many
binary, triple, and higher multiplicity stellar systems.
\citet{2014AJ....147...87T} studied the multiplicity of multi-body
systems within 67 pc of the Sun and found that 13\% of the
stellar systems have three or more stars. \citet{2010ApJS..190....1R} estimated that as much as 
8\% are three-body
systems, and~\citet{2013ApJ...768...33R}, using {\sc Kepler} data,
estimated that 20\% of close binaries have tertiary companions.
The {\sc Kepler} mission alone has discovered more
than 220 triple systems. In addition to the large number
of triple-star systems, a number of higher-order multiple
star systems have also been discovered by the {\sc Kepler}
mission and ground-based observations~\citep[][]{2015MNRAS.448..946B}. 
For example, \citet{2016MNRAS.462.1812R} have discovered
a quintuple star system that contains two small
eclipsing binaries orbiting each other. Each of the small
binaries have an orbital radius of 2 au, one of which
has a third star. The large binary formed by the small binaries
has an orbital radius of 12 au. Therefore, in such
multi-star systems, it is very likely that triple systems can
also be formed between a stellar black hole binary and
a nearby Sun-like star, or even in higher-order star systems~\citep{2012ApJ...748..105K}. 
This is more likely to occur in dense stellar environments such as the core of the
Milky Way or in the cores of globular clusters.
Nevertheless, as the amplitudes of oscillations for main-sequence stars
are a few orders of magnitude smaller than the amplitudes of red-giant stars,
and because the brightness of main-sequence stars are much smaller, consequently 
in order to observe the impact of gravitational radiation in main-sequence stars 
it will be more efficient to look for possible targets within the solar neighborhood.
Alternatively, because red-giant stars are much more bright,  it should be possible to look for potential red-giant 
star targets at much larger distances. Nevertheless, as we will discuss in this work, the most important  limitation on detection 
of the impact of gravitational waves on the pulsation spectrum  of Sun-like stars  
is related to the magnitude of gravitational radiation striking the star.

\smallskip
The interest in investigating the excitation of stellar oscillation modes by gravitational-wave 
radiation started in the 1960’s, when Joseph Weber published the first article about
the experimental detection of gravitational waves~\citep[i.e.,][]{1961grgw.book.....W}.  
Use of the Earth was proposed as a laboratory mass to
detect gravitational radiation~\citep{1961Natur.189..473F}. A few
years later,~\citet{1969ApJ...156..529D} studied the response of the Sun
and Earth to incoming gravitational radiation. Other publications have followed 
~\citep[i.e.,][]{1984ApJ...286..387B,2011ApJ...729..137S,2014MNRAS.445L..74M}. 
With the advance of the observations, this idea was pursued with the
gravitational techniques of helioseismology and
asteroseismology, either by studying the impact of
gravitational radiation on solar oscillations~\citep[i.e.,][and references therein]{2014ApJ...794...32L}  
or by extending this approach to other stars ~\citep[i.e.,][and references therein]{2015ApJ...807..135L}. 

\smallskip
In this article, we predict that stellar black hole binaries located
at short distances from Sun-like stars will excite a few of
their non-radial oscillations well above the photometric amplitude
variations caused by stochastic excitation.

We review some of the more current channels for the
formation of black hole binaries, and we discuss the
dynamic evolution of black hole binaries during the inspiral
phase. Finally, we evaluate the excitation of stellar
oscillations by gravitational waves.

\section{Formation of Stellar Black Hole Binaries}
\label{sec-2}

The recent discovery of the GW150914~\citep[][]{2016PhRvL.116f1102A} 
and the GW151226~\citep[][]{2016PhRvL.116x1103A} black hole
binaries~\footnote{GW150914: located at a distance of $410\; {\rm Mpc} $,
with black hole masses of $36\; M_\odot$ and $29\; M_\odot$; and
GW151226: located at a distance of $440\; {\rm Mpc} $, with
black hole  masses of  $14.2\; M_\odot$ and $7.5\; M_\odot$.} 
is an indisputable confirmation that massive stellar
black hole binaries are very likely quite common. Such compact
binary systems can form, either from isolated binary stellar
systems in galactic fields, or as a result of dynamic interactions
of stars in dense stellar environments such as the cores of
stellar clusters or galaxies.

\smallskip
In the first case, known as a standard scenario, the formation
of black hole binaries follows through a complex number of
stages~\citep[i.e.,][]{2014LRR....17....3P}. Firstly, two massive
stars form in a common stellar envelope, and a stable mass
transfer is established between the two stars. By the action of
gravity, the core of the largest star collapses into a black hole.
Then, a second mass transfer starts from the external layers
of the remaining star to the black hole. During this phase
the binary continuously loses mass and angular momentum.
The compact binary forms when the core of the second star
also collapses into a black hole~\citep{2005ApJ...632.1035O}.
Although there are several stellar models to explain the formation
of black hole binaries, the vast majority follow this
evolutionary path~\citep{2008ApJS..174..223B,2002ApJ...572..407B}. For instance,~\citet[][]{2016MNRAS.458.2634M,2016A&A...588A..50M} 
recently proposed that massive stars, which throughout their lifetimes remain in a close compact
binary, maintaining a rapid rotation and a chemically homogeneous
interior, rapidly end by forming a massive stellar black hole
binary.

\smallskip
In the second case, the black hole binaries form in massive
stellar clusters, due to the drift of single or binary black holes
to the core of the cluster by dynamical friction~\citep{2016ApJ...824L...8R}.
 In these dense stellar environments, the gravitational interaction 
of stars and black holes leads to the formation
of multi-stellar binary systems, some of which are quite
often ejected from the cluster~\citep{1993Natur.364..423S}.
Such interactions preferentially keep the heaviest objects in
binaries and eject the lightest ones leading to a concentration
of heavier black hole binaries in the cluster core~\citep[]{1975MNRAS.173..729H}.   
In some cases, such processes leads to the formation
of triple compact star systems. \citet[]{2016arXiv160402148M} suggest
that the presence of a nearby third star can be determined by
the perturbation caused to the gravitational wave emitted
during the merger of the two black holes of the inner binary.

\smallskip 
It is well accepted that both evolution scenarios predict the
formation of black hole binaries with masses comparable to
those recently discovered~\citep{2016ApJ...818L..22A,2016PhRvL.116x1103A},  
though further improvements are required to fine-tune their
evolutionary paths. Nevertheless, all scenarios predict that the
more massive binaries are formed in stellar environments of
low metallicity~\citep[e.g.,][]{2013ApJ...779...72D}. Equally, the formation
of binaries by dynamical friction is intensified in the
cores of globular clusters, where there is a high density of
stars with low metallicity~\citep{2016PhRvD..93h4029R}. 

\smallskip
One test of all these formation scenarios is that the black
holes of such binaries must merge on a time-scale smaller
than the age of the universe ($\sim$ 13.7 Gyr). For this to happen,
the orbital separation of the black holes must be sufficiently
small to allow rapid binary contraction by energy loss to
gravitational radiation, culminating in the coalescence of the
two black holes. Although this property is verified for all
binaries, the binary formation time-scales and their
coalescence time-scales are very different. For instance, in the
standard scenario, compact binaries form on a time-scale of
10 Gyr~\citep{2014LRR....17....3P}. In a variation of this
scenario, a close binary formed by two massive stars
coalesces into a black hole binary with a total mass of 80 $M_\odot$
in a record time of 2.6 Gyr~\citep{2016ApJ...824L...8R}. In this
case, black hole binaries created by stellar encounters in
globular clusters have a typical waiting time to coalescence of
6.3 Gyr.

\section{Stellar Black Hole Binaries in the Inspiral Phase}
\label{sec-3} 
The final evolutionary stage of a black hole binary, as
recently validated by the discovery of GW150914, follows a
well-defined process of three phases~\citep{2014LRR....17....2B,2000PhRvD..62f4015B}: it begins with the black
holes being far apart and rotating in near-perfect circular
orbits (inspiral), entering into a stage of rapid orbital
shrinking (speed-up), and terminating by merging into a
single black hole (ringdown).  Here our study
focuses on an approximate description of the first two phases
in the sprit of~\citet{2016arXiv160706818D}.

\smallskip
We start by assuming that the two black holes in the binary
are in a circular orbit. The detector star is considered to be
sufficiently far away from the binary, such that the incoming
gravitational radiation is described by a plane wave field, but
near enough that the redshift corrections are ignored in the
gravitational-wave spectrum. During the two initial phases of
the binary contraction, the frequency and amplitude of 
gravitational-wave strain vary as follows.

\smallskip
(a) As the gravitational frequency $\nu$ is two times the orbital
frequency, Kepler's laws show that $\nu$ increases with the
reduction of the orbital radius of the binary. As usual,
we define the angular frequency $\omega$ to be equal to $2\pi\nu$. The
smaller black hole slowly enters in an adiabatic inspiral process,
by going through a succession of quasi-circular orbits
during which it loses energy by gravitational radiation. Consequently,
the time to coalescence $\tau$  decreases as $\nu$ increases,
\begin{eqnarray} 
\nu/\nu_c=0.0728\left({\nu_c\tau}\right)^{-3/8} 
\label{eq:ftau} 
\end{eqnarray}
where $\nu_c$ is a characteristic frequency of the binary system. 
$\nu_c$ is equal to ($c^3/(GM_c)$), where $c$ and $G$ are the speed of light and Newton constant, 
and  $M_c$ is the chirp mass of the binary system, $M_c\equiv(m_1 m_2)^{3/5}/M_t^{1/5}$, where $m_1$ and $m_2$ 
are the masses of the two black holes and $M_t=m_1+m_2$.  
Similarly, the orbital radius $a$, which is equal to $(GM_t/(\pi \nu))^{1/3}$,  decreases as $\nu$ increases.
\begin{figure}
\centering
\includegraphics[scale=0.4]{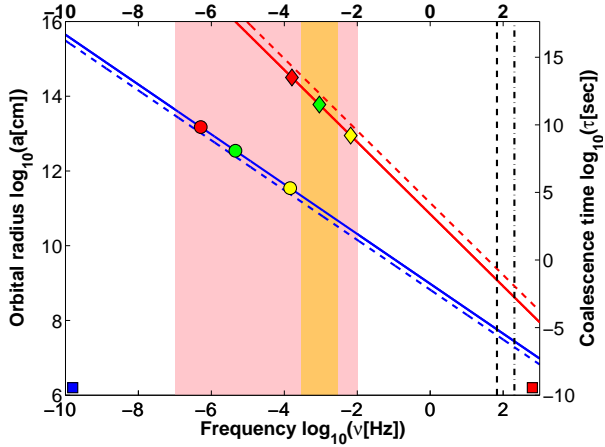} 
\caption{
Frequency time variation of the gravitational wave emitted by the GW150914 (continuous curves)
and GW151226 (dashed curves)  black hole binaries:
the blue (left axis) and red (right axis) lines correspond to the variation of the orbital radius 
($a[{\rm cm}]$) and time to coalescence ($\tau [{\rm sec}]$) as the binary proceeds toward coalescence.  
The vertical black lines correspond to the frequency $\nu_{\rm max}$, 
for GW150914 (dashed line) and GW151226 (dotted-dashed line). 
The pink area defines the frequency range of non-radial oscillations observed in Sun-like stars: 
main-sequence and sub-giant stars from $10^{-4} {\rm Hz}$ to 
$ 10^{-2} {\rm Hz}$~\citep{2008ApJ...682.1370K}; and red-giant stars from 
$10^{-7} {\rm Hz}$ to $10^{-3} {\rm Hz}$~\citep{2013sf2a.conf...25M}.
The orange area corresponds to the frequency oscillation range of the star detector shown in Figure~\ref{fig:2}.
The red, green and yellow circles correspond to the orbital radius $1$ au, $50\;R_\odot$ and $5\;R_\odot$ solar radius.   
The red, green and yellow diamonds correspond to the time to coalescence of  $10^{6}$ year, $10^{4}$ year and $50$ year, respectively.   
}
\label{fig:1}
\end{figure}

The inspiral phase ends when the radial distance between
the two black holes is shorter than the last stable circular orbit,
also known as the innermost stable circular orbit (ISCO).
The frequency of this orbit $\nu_{\rm ISCO}$ is approximately equal to
$2.2 kHz\; (M_\odot/M_t) $. When this orbit is passed, the two black
holes merge and coalesce. Hence, $\nu_{\rm max}$, the frequency of the
gravitational wave emitted on the ISCO orbit is equal to twice
the $\nu_{ISCO}$~\citep{Maggiore:2008tk}. $\nu_{\rm max}$ corresponds also to the
maximum frequency of the gravitational radiation emitted in
the inspiral phase.

\smallskip 
(b) The strain $h$ increases as the binary system approaches  coalescence. 
The nature of gravitational radiation determines that $h$ has two polarized components,  
$h_{+}$ (plus polarization) and $h_{\times}$ (cross polarization). These quantities are 
written in a condensed form: from the start of the inspiral phase until coalescence, 
$h_{k}(\tau)$ is equal to $ {\cal A}_\star\; g_{k}(\varphi)\; {\rm C}_{k} [\Phi\left(\tau\right)]$  
where $k$ is equal to  $+$ or $\times$, ${\cal A}_\star$ is the amplitude, $g_{k}$ are geometrical functions 
depending of the directional angle $\varphi$, 
and  ${\rm C}_{k} $ are circular functions depending of the phase $\Phi\left(\tau\right)$~\citep{Maggiore:2008tk}.
The strain amplitude ${\cal A}_\star$ reads
\begin{eqnarray}
{\cal A}_\star =c/(d_\star \nu_c)\;\left(5/(\nu_c\tau)\right)^{1/4},
\label{eq:htk} 
\end{eqnarray} 
where $d_\star$ is the distance of the black hole binary system to the  detector star.

\medskip
Figure~\ref{fig:1} shows the time variation of two binary parameters 
-- orbital radius and time to coalescence, as a function
of $\nu$  for the two recently discovered black hole binaries:
GW150914~\citep{2016PhRvL.116f1102A}  and GW151226~\citep{2016PhRvL.116x1103A}. 
Since their formation (the upper limit is the age of
the universe), these binaries have emitted gravitational radiation in the
frequency range of of $10^{-7}$ -- $10^{-2}$ Hz.  During this time, the orbital
radius reduces to a tenth of the solar radius ($\sim 10^{10}\;{\rm cm}$), 
being only 50 years ($ \sim 1.5\; 10^9\;{\rm sec}$) away from the binary coalescence.
As illustrated in Figure~\ref{fig:1} during most of the inspiral
phase, the gravitational radiation emitted can stimulate
non-radial oscillations of Sun-like stars located nearby these
binaries.
          

\begin{figure}
\medskip
\centering
\includegraphics[scale=0.40]{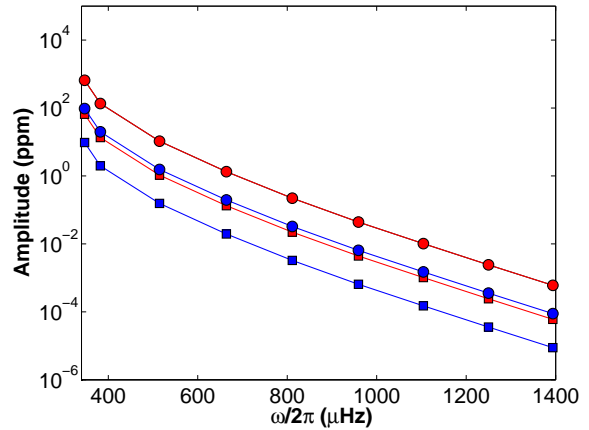}
\caption{Maximum intensity amplitude of quadrupole modes
computed for a solar main-sequence star excited by a nearby 
GW150914 (red curve) and GW151226 (blue curve)  black hole binary. 
The detector star is located at a distance of 10 au (square symbol) or  1 au (circle symbol).
The orbital radius of these binaries varies from $3\;R_\odot$ to $0.66\;R_\odot$,
during which the frequency of the gravitational waves emitted varies from $3\;10^{-4}\;{\rm Hz}$ to 
$3\;10^{-3}\;{\rm Hz}$ (corresponds to $log_{10}(\nu[Hz])$ varies from -3.4 to -2.8,
 see the orange area in Figure~\ref{fig:1}).}
\label{fig:2}
\end{figure}
 
\section{Impact of gravitational waves in non-radial oscillating stars}
\label{sec-4}
The response of the Sun to resonant gravitational radiation has
been discussed by several authors, following the detailed study
made originally by~\citet{1984ApJ...286..387B}. Most aspects
of the analysis for other stars are similar. We follow the
standard discussion about gravitational radiation emission~\citep{Maggiore:2008tk}.

\smallskip
Here, to compute the stimulus of non-radial oscillations of
a Sun-like star struck by gravitational waves from a nearby
black hole binary, we follow the study and notation of~\citet[][and references therein]{2015ApJ...807..135L}.
Accordingly, the stimulation of gravitational waves only excites stellar modes of
even degree $\ell$. Of these, we focus our study on the quadrupole
modes ($\ell=2$), for which the gravitational forcing is the
greatest. Because our goal is to compute the amplitude of these
modes, the problem is simplified by focusing our attention
on the quadrupole acoustic modes of frequency $\omega_n$ and order $n$
(overtone). 

\smallskip
In general, because a black hole binary radiates gravitational
waves, it loses energy and thus spins up. This process is
accompanied by a reduction of the binary size and an increase
of the orbital frequency. 
Accordingly to \citet{2014ApJ...794...32L}, 
we have computed the maximum impact of gravitational waves on photospheric velocity amplitudes of resonant stellar quadrupole modes; for convenience, we choose here to write as the variation of photometric amplitudes
\begin{eqnarray} 
\left(\frac{A_n}{{\rm ppm }}\right)=
\frac{A_\odot}{{\rm v}_\odot}\;f_s\;\frac{{\cal A}_\star}{{\rm cm s^{-1}}}\; \frac{L_n}{\alpha_s}\frac{\omega_n^2}{\eta_n},
\label{eq:Aphoto}
\end{eqnarray} 
where  ${\cal A}_\star$ is the strain amplitude of the gravitational wave (see equation~\ref{eq:htk}), 
and $\omega_n$ and $\eta_n$ are the angular frequency and damping rate of the star's mode.  $\alpha_s$ is a constant equal to $2\sqrt{2}$, for which the exact value is  fixed by observations.
$f_s$ is equal to $\sqrt{T_{\odot}/ T_{\star}}$, where $T_{\odot}$ and $T_{\star}$  are the effective
temperatures of the Sun and the star.  $A_\odot$  is the  maximum observed bolometric amplitude
and  ${\rm v}_{\odot}$ is the maximum surface velocity of the mode in the Sun's surface. The values of $A_\odot$, ${\rm v}_{\odot}$ and $T_\odot$ are equal to
$2.53\pm 0.11$ ppm  (parts-per-million), $ 18.5 \pm 1.5 {\rm cm\; s^{-1}}$ and $5777 ^oK$.  
All the previous quantities are evaluated at the solar photosphere~\citep{2009AA...495..979M}. 
In the  determination of $A_n$ (equation~\ref{eq:Aphoto}), we use the 	
photospheric velocity amplitude expression $V_n={\cal A}_\star L_n/\alpha_s\;\omega_n^2/\eta_n$ from~\citet{2014ApJ...794...32L}.
The conversion from photospheric velocity  to photometric amplitude variations is made using the relation obtained  by~\citet{2009AA...495..979M}. 
We note that the photometric amplitude given by equation (\ref{eq:Aphoto})  can be easily modified for stars in other 
stages of   evolution. In particular, for red-giant stars, 
this can be done by using  a scale relation like the ones obtained
 by~\citet{2011ApJ...743..143H} and~\citet{2011A&A...531A.124B}. These scale relations were obtained empirically from large data photometric sets by the {\sc Kepler} Mission. These observational data sets comprise photometric time series for stars in quite distinct stages of evolution, like  the main-sequence  and red-giant phases.
$L_n$ is the effective modal length that depends on the properties of the quadrupole mode of order $n$: $L_{n}=1/2\; R_\star\; |\chi_{n}|$,  where $R_\star$ is the star's radius and $\chi_{n}$ is a coefficient that depends on the eigenfunction  $\xi_n$: 
\begin{eqnarray}
\chi_{n}=\frac{3}{4\pi \bar{\rho}_{\star}}
\int_0^1 \rho (r)\left[ \xi_{r,n}(r)+3 \xi_{h,n}(r) \right] r^3 dr.
\label{eq:chin}
\end{eqnarray}  
where $\rho$ and $\bar{\rho}_{\star}$ are the density profile and mean density of the star, 
and $\xi_{r,n}$ and  $\xi_{h,n}$ are the radial and horizontal components $\xi_n$~\citep{2014ApJ...794...32L}. 

\medskip
If the strain of the incoming gravitational wave is considered to
be constant, equation (\ref{eq:Aphoto})  becomes identical 
to the previous amplitude predictions (for generic gravitational-wave sources), like the ones obtained by~\citet{2011ApJ...729..137S} 
and~\citet{2014ApJ...794...32L}. Equation (\ref{eq:Aphoto}) differs from the 
previous ones  because in this  case it takes into account
 the variation of $A_n$ due to the orbital contraction of the black hole  binary
(see equation~\ref{eq:htk}).
Accordingly,  from the previous equations
(\ref{eq:htk}) and (\ref{eq:Aphoto}), the new expression reads
\begin{eqnarray} 
\left(\frac{A_n}{{\rm ppm }}\right)=
\frac{A_\odot}{{\rm v}_\odot}\;f_s 
\frac{h_\star}{{\rm cm s^{-1}}}
\;\left(\frac{5}{\nu_c\tau}\right)^{1/4}
\frac{L_n}{\alpha_s}\frac{\omega_n^2}{\eta_n},
\label{eq:Aphoto2}
\end{eqnarray} 
where  $h_\star=c/(d_\star \nu_c)$. The above expression is valid provided that the frequency of the gravitational wave is in resonance with the frequency of the stellar mode and its variation can be neglected. 
 This equation includes the term ${A_\odot}/{{\rm v}_\odot}f_s\times  
({h_\star}/{{\rm cm s^{-1}}})\times \left({5}{(\nu_c\tau)}\right)^{1/4}$
that converts velocity amplitudes to photometric amplitudes, and also takes into account the  strain amplitude time variation emitted by the compact binary.

\medskip
Figure~\ref{fig:2} shows the photometric amplitude variations
of quadrupole acoustic modes for resonant gravitational
waves emitted by GW150914 
and GW151226  black hole binaries. In this
calculation, the values of $\omega_n$ and $\chi_{n}$  are obtained for the present Sun~\citep[i.e.,][]{2014ApJ...794...32L,2012RAA....12.1107T}, and 
$\eta_n$ corresponds to damping rates for realistic solar envelope models~\citep{1999A&A...351..582H,2005A&A...434.1055G}.
The result shows that these black hole binaries
when in the vicinity of typical Sun-like stars of a solar
mass are able to strongly excite the lowest order overtones
of these modes, producing photometric amplitude variations
well above $100$ ppm. In some extreme cases, the amplitude
variation is above $10^3$ ppm.
These values are much larger than the photometric amplitudes of just a few ppm’s observed for
low-degree acoustic modes in main-sequence stars~\citep[][]{2011A&A...534A...6C},
as predicted to be stochastically excited by turbulent
convection~\citep{1999A&A...351..582H}. 
In a recent study,~\citet{2011ApJ...743..143H} found that main-sequence and
red-giant stars show photometric amplitudes increasing from a few ppm up to 1000 ppm.

\medskip
Among the possible physical processes capable of suppressing   
the impact of a gravitational wave in a stellar mode (see equation~\ref{eq:Aphoto2}), the damping rate $\eta_n$ is the most significant one. It is true that until now, no low-order quadrupole modes have been successfully observed in Sun-like stars,  possibly because the fluid motions of the upper layers of stars are suppressing the stochastic excitation of the modes due to turbulence in the layers beneath. Indeed, the energy exchange  between convection pulsations could affect the  amplitude of such modes. This specific point was discussed by several authors, among others including~\citet{1997MNRAS.288..623C} and~\citet{2006ESASP.624E..97D}, among others. 
Nevertheless, we note that the excitation provided by gravitational radiation is external to the star and therefore the prediction of the mode
excitation is simpler that  the case of modes excited by the turbulent processes of the convection zone, which depends of many local physics processes  
such as magnetic fields and differential rotation. We believe that the predictions made in this study  
 are realistic since the theoretical damping rates used in this study follow from the same theory of stellar pulsation that successfully predicted the excitation and damping of stellar oscillations for the Sun and many Sun-like
stars~\citep{1999A&A...351..582H,2005A&A...434.1055G,2012RAA....12.1107T,
2013ASPC..479...61B}.  This theory successfully describes the amplitude of
the observed acoustic modes of low-degree $\ell$ and high-order $n$.  
Therefore, it is reasonable to assume that at lower frequencies such a theoretical model of stellar pulsations is still valid. 
On the unlikely possibility that  lower-order modes have damping rates 
comparable to the observed value  of $10^{-3}\;\mu {\rm Hz}$ (and predicted by theoretical pulsation models) for stellar modes  with  $\omega/2\pi \sim 1000\;\mu {\rm Hz}$ the amplitude of such modes will be below the {\sc Plato} threshold 
of detection. Nevertheless, as explained by~\citet{2014ApJ...794...32L} 
the current theoretical predictions suggest that such stellar modes with 
$\omega/2\pi < 1000\;\mu {\rm Hz}$ have much smaller values of damping rates.  

\medskip
The next generation of satellites, like {\sc Plato}~\citep{2013EGUGA..15.2581R}
should be able to measure photometric amplitudes with the
necessary precision to detect such an impact of gravitational
waves in low-order acoustic quadrupole modes. The current
estimate for the {\sc Plato} mission predicts that the telescope
will be able to measure photometric amplitude variations
above 3.2 ppm after 5 days (or 1.3 ppm after 1 month)
of continuous observation.


\section{Summary and Conclusions}
\label{sec-5}

The recent detection of gravitational waves from two distant
black hole binaries of intermediate mass implies that such binary
systems are quite common in the universe. This also
hints at the existence of very efficient mechanisms for the formation
of these binaries. In this work, we mention two of
these processes: as the end product of the evolution of binaries
of massive stars and as resulting from the gravitational
encounters of stars in high density stellar environments, as
found in the cores of stellar clusters. Moreover, the formation
of black hole binaries from progenitor stellar binaries is facilitated
in low-metalicity stellar environments~\citep[e.g.,][]{2016ApJ...818L..22A}, 
most notably in the cores of globular clusters~\citep{2016PhRvD..93h4029R}.  
These facts substantiate the existence
of regions in the Milky Way with a mixed population of
stellar black hole binaries and Sun-like stars.

\smallskip
This study suggests that the quadrupole acoustic modes
of stars located nearby black hole binaries in some cases can  have relatively large amplitudes well above the experimental threshold of the current  asteroseismic satellites.  Nevertheless, there are important caveats that need to be addressed if one wishes to successfully find  such stars.  In particular, these amplitude predictions do not take into  account the visibility of stellar modes, which depends on the photometric properties of the star, as well as the degree and azimuthal order of the mode. This important point was discussed in the context of gravitational
radiation for the first time by~\citet{2014ApJ...784...88S,2015ApJ...810...84S} for the case of high-precision radial velocity of solar gravity modes. 
These authors have shown that  visibility effects lifted the degeneracy in the azimuthal order of the mode leading to the reduction of amplitudes  by a few orders of magnitude. However, as pointed out by~\citet{1990A&A...227..563B}, these visibility corrections depend on the properties of stellar modes and on the detailed structure of the upper layers and atmosphere of the star. Unfortunately, these stellar regions are still poorly described by the current stellar models~\citep[i.e.,][and references therein]{2013MNRAS.435.2109L}. Nevertheless, this difficulty should be overcome since the large quantity and high-quality data made available by the current astereoseismic satellites should allow astronomers to
accurately model the upper layers of Sun-like stars in the near future.

\medskip
The possibility of observing the imprint of gravitational radiation on the stellar pulsation spectrum of a Sun-like star by a future asteroseismology mission 
is limited by the existence of many unknowns related to the dynamic properties of the stellar systems being observed, as well as by the experimental specifications of the telescopes being used for the observations.  Nevertheless, there are a few points that we can make that could help the discussion to successfully discover such stars.    

\smallskip
%
%
Since it is now known that stars show Sun-like oscillations in many stages of the stellar evolution including main-sequence, sub-giant and red-giant phases. 
It is possible that some of these stars are located sufficiently near such black hole 
stellar systems.  As discussed previously, black hole binaries can form in two types of scenarios: in stellar multiple systems where the black holes form as the end stage of stellar evolution, or the binaries are created due to the dynamical interactions between stars. Both processes are more likely to occur in dense stellar regions of the Milky Way,  like the ones found in the core of stellar clusters and the regions of intense stellar formation near the galactic center. 
In any case, for the spectrum of a Sun-like star to be affected by gravitational radiation, it is paramount that the black hole binary be located relatively near the Sun-like star, possibly for even the black hole binary and the Sun-like star to be part of the same multiple stellar system~\citep{2016ComAC...3....6T}.  There is strong evidence to support that a significant fraction of stars resides in a multi-stellar system including binaries~\citep{2010ApJS..190....1R}.  \citet{2015MNRAS.448.1761W} estimate that 54\% of Sun-like are in binaries  or multiple stellar systems. \citet{2013ApJ...768...33R}, using a  limited range of orbital periods for the triple-star systems,  estimate that at least 20\% of all close binaries have tertiary companions.
Recently,~\citet{2015AJ....150..130R} have observed a few triple systems with solar-type stars in the optical as well as in the near-infrared~\citep{2017AJ....153..100R}
and~\citet{2017MNRAS.tmp..145R} discovered a quadruple stellar system
and quintuple stellar system~\citep{2016MNRAS.462.1812R}.
Moreover, as much as 70\% of massive stars are observed to have  companions~\citep{2007ApJ...670..747K}. This is significant, since some of these massive stars could lead to the formation of black holes, in which the instantaneous supernova expulsion shell will not affect the companion stars~\citep{2016ComAC...3....6T}. However, the explosion will modify the dynamics of the multiple stellar system~\citep{2012MNRAS.424.2914P}.

\smallskip
%
%
In spite of not knowing how many stellar black hole binaries exist in the Milk Way that are located nearby Sun-like stars, at least we can make an estimation of the total number of stellar black hole binaries in our Galaxy.  
Recently,~\citet{2017arXiv170101736C} using the merger rates of 
LIGO found that  the number of black hole binaries in the Milky Way  $N$ 
with semi-major axis $a$ smaller than $a_{bh}$  to be given by $N(a \le a_{bh})\approx K_{bh} R_{100} \left(a_{bh}/10R_\odot\right)^4$ where $K_{bh}$ is a rate factor related to the black hole binary mass  and  $R_{100}$ is the rate in units of $100\; {\rm Gpc}^{-3}{\rm yr}^{-1}$.   If only binaries with black hole mass of $30\; M_\odot$ are take into account,  then $K_{bh}=3\times 10^{2}$. However, if the same calculation is repeated assuming a power law for the mass function of massive stars, i.e., taking into account all  binaries (with black holes of different masses) with the same orbital radius, then $K_{bh}=3\times 10^{4}$. For instance,
$N$ is equal to 20 in the case of binaries formed by two black holes with a mass of  $30\;M_\odot$ each and with a semi-major $a$ axis smaller than 
$a_{bh}\sim 5\; R_\odot$. 
In the case in which the black hole masses in the binary can take other values,
$N$  will increase to $2000$. The same expression predicts the 
existence of $19\times 10^6$  black hole binary systems with  $a_{bh}\sim 50\; R_\odot$. This type of binaries will emit gravitational radiation that will
excite the quadrupole modes of nearby Sun-like stars, black hole binaries with a $a_{bh}$ $\sim 50\; R_\odot$  and $\sim 5\; R_\odot$  will emit waves  with a frequency greater than  $10^{-6}\;{H z}$ and $10^{-4}\;{H z}$ (see figure~\ref{fig:1}).

\smallskip
%
%
%
%
%
The important point is that the Sun-like stars to be observed by the asteroseismic satellite  must be located at a short distance of the black hole binary. Therefore,
we will increase the chances to make such detections if the telescope is able to observe stars at large distances. Among the star targets observe by  the {\sc Corot} and {\sc Kepler} satellites~\citep{2010AA...517A..22M,2011ApJ...743..143H}, 
a few tens of thousands of stars have been discovered  to show Sun-like oscillations. 

\smallskip
The original {\sc Kepler} mission alone has made such discoveries only 
by looking   in  a fixed direction   of the sky; however, with its renew observational program K2 mission~\citep{2014ApJS..211....2H}, the survey is now extended to many other directions of the Milky Way, increasing significantly the chances to look for Sun-like stars near black hole binaries.  Although there is a possibility that such phenomena will be found within the current asteroseismic data sets,  it is more likely that it will happen in a future mission, such as {\sc Plato}~\citep{2014ExA....38..249R} and {\sc Tess}~\citep{2014SPIE.9143E..20R} satellites, or even more likely  that it will happen in a post-{\sc Plato} asteroseismology mission for which specific observational targets could be chosen  with this objective in mind.  To succeed in this endeavour, it is necessary to understand which are the most likely stellar environments where it could be found the targets with Sun-like stars nearby black hole binaries could be found.

\smallskip
The  {\sc Plato} satellite is expected to obtain 1,000,000 stellar photometric light curves for stars with a  magnitude smaller 14 over the course of the full mission~\citep{2014ExA....38..249R}.    If we consider  that a proportional of 10\%  of stars have Sun-like oscillations like it was found in the case of the {\sc Kepler} mission, a total of 100,000 should also have Sun-like oscillations.  
In principle, this mission should be able to observe a typical star like the  sun up  to a distance of 2 kpc  and a red-giant star up to a distance of 20 kpc, if it was to the light obstruction by dust,  the {\sc Plato} mission could observe red-giant stars in the center of the Milk Way (at a distance of 8 kpc).   Therefore, it is reasonable to expect that some of these Sun-like stars are located near the black hole binaries. 

\smallskip
Since some Sun-like stars that likely are located nearby black hole binary systems can be found in the dense stellar regions of the Milky Way obscured by dust, an asteroseismology mission dedicated to observe in the near-infrared band could be an excellent alternative to optical asteroseismology.  As  the  star brightness and amplitude of stellar modes in this band is only a  factor of five smaller than the optical measurements counterpart, this could significantly increase the possibility of  finding such types of stars once it allows us to observe in the dense regions of the Milky Way. In spite of the asteroseismology of the near-infrared being very poorly understood
at present,  this technique could be particularly useful to look for stars in the dense regions of the galactic disk in the direction of the galactic center, where it is known to have large populations of compact stars.

\smallskip

We show here that if the generally accepted ideas about
general relativity are correct, the low-order overtones of quadrupole acoustic
modes of Sun-like stars could be driven by gravitational
radiation of nearby stellar black hole binaries,
identical to the ones discovered by the Advanced LIGO collaboration~\citep{2016PhRvL.116f1102A,2016PhRvL.116x1103A}.
Nonetheless, only if there is a sustained resonance between the
incident gravitational radiation and the stellar quadrupole mode can
there be a chance of an observable response from the star. In
general, the energy loss by gravitational radiation causes the
black hole binary to spin up, prohibiting the gravitational wave from remaining in resonance with a
mode for a period  long enough to have a perceptible effect.
However, because the orbital frequency drift is
negligible for many of these black hole binaries during
most of the inspiral phase,
oscillations in a solar mass star can be stimulated to an
observable photometric amplitude by a binary  at a
distance of 1 au. In particular, we predict that the
photometric amplitude variations of these modes
are well above the experimental threshold of
present and future asteroseismic satellite missions
~\citep{2013EGUGA..15.2581R}. Therefore, it is even possible that
these observations could have already been made.

\begin{acknowledgments}
The author thanks the anonymous referees for the insightful comments and suggestions. In particular, the suggestion to include a discussion 
about the possibility of observing such phenomena by a future astereoseismology
satellite was very helpful. 	
The work of I.L. was supported by grants from "Funda\c c\~ao para a Ci\^encia e Tecnologia"  and "Funda\c c\~ao Calouste Gulbenkian". 
We are grateful to the authors of ADIPLS and CESAM codes for having made their codes publicly available.
The work of J.S. was supported by ERC Project No. 267117 (DARK) hosted by  Universit\'e Pierre et Marie Curie (UPMC) -- Paris 6.
\end{acknowledgments}
  
\bibliographystyle{yahapj}

 
\end{document}